\documentclass[12pt]{article}
\usepackage{natbib}
\usepackage{color}
\usepackage{amssymb,bm,amsmath,physics,amsthm}
\usepackage{graphicx}

\usepackage{booktabs}
\usepackage[margin=1in,lines=25]{geometry}
\usepackage{setspace}
\newtheorem{proposition}{Proposition}
\renewcommand{\hat}{\widehat}

\begin{document}

\title{Spatial Weibull Regression with Multivariate Log Gamma Process and Its
Applications to China Earthquake Economic Loss
}


\author{Hou-Cheng Yang\thanks{Department of Statistics, Florida State
University}~~~~Lijiang Geng\thanks{Department of Statistics, University of
Connecticut}~~~~
Yishu Xue$^\dagger$\thanks{yishu.xue@uconn.edu}~~~~
        Guanyu Hu$^\dagger$
}

%
\date{}

\maketitle

\begin{abstract}
Bayesian spatial modeling of heavy-tailed distributions has become increasingly
popular in various areas of science in recent decades. We propose a Weibull
regression model with spatial random effects for analyzing extreme economic
loss. Model estimation is facilitated by a computationally efficient Bayesian
sampling algorithm utilizing the multivariate Log-Gamma distribution.
Simulation studies are carried out to demonstrate better empirical performances
of the proposed model than the generalized linear mixed effects model.
An earthquake data obtained from Yunnan Seismological Bureau, China is
analyzed. Logarithm of the Pseudo-marginal likelihood values are
obtained to select
the optimal model, and Value-at-risk, expected shortfall, and
tail-value-at-risk based on posterior predictive distribution of the
optimal model are calculated under different confidence levels.
\end{abstract}
\section{Introduction}\label{sec:intro}

Extreme geological disasters often cause catastrophic impact to both
environmental systems and human society. For example, an earthquake can cause
ground shaking, ground rupture, landslides, tsunami, etc. All such effects pose
serious threats to the environment as well as humans,
and cause tremendous economic losses and casualties.
Study of the influential factors for, and consequences of such extreme
environmental events, is of great value to both the environment and the human
society.

Both the locations and outcomes of extreme events received attention. As the
locations of earthquakes are often random realizations of an underlying process,
which can be related to various geological factors, spatial point process models
have been developed to capture patterns in such locations
\citep{vere1970stochastic,ogata1988,schoenberg2003,hu2019bayesianmodel}. 
Regression-based models have been used to analyze factors
that influence the outcomes, e.g., earthquake magnitudes
\citep{charpentier2015modeling,hu2018bayesian,yang2019bayesian,xue2019online}.
The economic loss incurred by such disastrous events are often studied using
extreme value theory \citep[EVT;][]{coles1996bayesian,bali2003extreme} in
different fields such as economics and finance, insurance, environmetrics, and
geology. The bridge between spatial factors and the economic losses due to
extreme events, however, have not been fully established. \cite{li2016bayesian}
proposed a Bayesian approach for a total of four mixture models to depict
catastrophic economic losses caused by earthquakes. Covariates and
spatial-dependent structures are, nevertheless, missing from the model, which
can be a major disadvantage as economic losses caused by earthquakes tend to be
spatially varying, and are highly correlated with certain predictors such as
magnitudes of earthquakes, or categories of hurricanes.

In this work, we propose a hierarchical Bayesian approach for analyzing economic
losses caused by extreme events. A spatial generalized linear mixed effects
model (GLMM) is proposed, where spatial random effects
\citep{carlin2014hierarchical} are incorporated into the model to allow for
information leveraging from neighbors. We choose to use the
traditionally-popular Weibull distribution to model economic loss because of its
heavy tail. The multivariate log-gamma distribution
\citep[MLG;][]{bradley2018computationally} is used as the prior for the Weibull
distrbution \citep{xu2019latent}. As MLG enjoys conjugacy, closed forms for
posterior distributions can be obtained, which facilitates efficient
computation. Three risk measures based on the posterior predictive distribution
are introduced. Our simulation studies show promising empirical performance of
the proposed Bayesian methods as the parameter estimation is fairly accurate.
The model is further illustrated with an earthquake dataset from Yunnan, China,
and it identifies impact factors that influence the final incurred loss.

The rest of article are organized as follows. Section~\ref{sec:motivate_data}
gives a brief introduction and description of the motivating data. In
Section~\ref{sec:method}, we develop the spatial Weibull regression model and
risk measures based on the posterior predictive distribution, and examine
theoretical properties of the proposed model. Furthermore, Bayesian model
selection criteria Logarithm of the Pseudo-marginal likelihood (LPML) is used
for model comparison in Section~\ref{sec:model_section}. In addition, extensive
simulation studies are conducted in Section~\ref{sec:simu} to investigate
empirical performance of the proposed model. In
Section~\ref{sec:real_date}, we implement our model using Chinese earthquakes
data from 1950 to 2014. Section~\ref{sec:discussion} concludes with a brief
discussion. For ease of exposition, all proofs are given in the
appendices.

\section{Motivating Data}\label{sec:motivate_data}

\begin{figure}[tbp]
	\centering
	\includegraphics[width=0.8\textwidth]{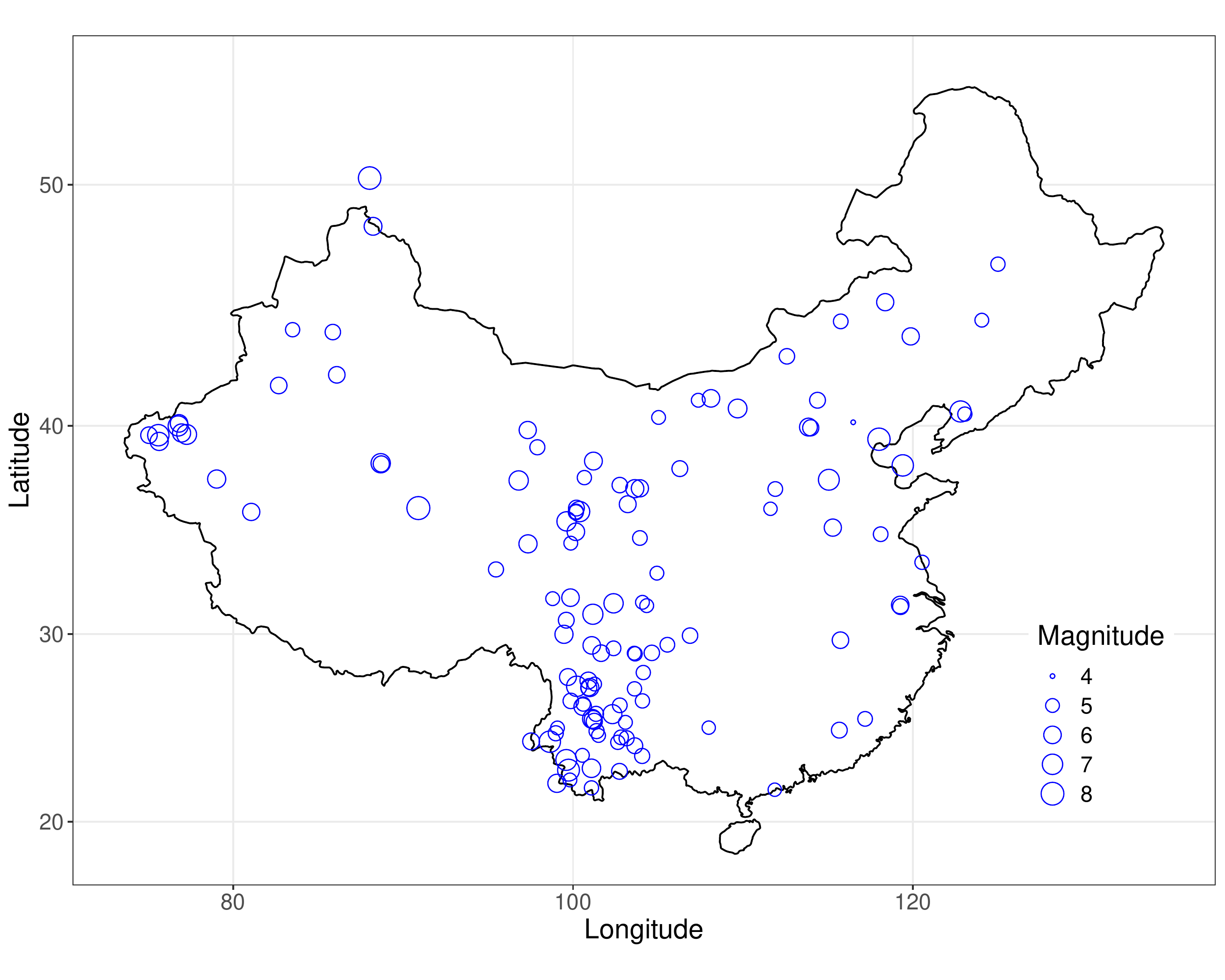}
	\caption{Visualization of earthquake locations in and close to mainland
China between 1950 and 2014. Larger circle indicate more severe economic
loss.}\label{fig:map}
\end{figure}

Similar to \citet{li2016bayesian}, we analyze the direct economic losses caused
by earthquakes which occurred in and close to mainland China between 1950 and
2014, collected by Yunnan Province Seismological Bureau. In this data set,
earthquake magnitude is a number that characterizes the relative size of the
earthquake, which is based on a measurement of the maximum motion recorded by a
seismograph. The location (latitude, longitude) is recorded for each occurrence.
An indicator variable denoting whether an earthquake occurred in urban or rural
areas is also present. A visualization of the earthquake locations and
magnitudes is shown in Figure~\ref{fig:map}.

\begin{figure}[tbp]
	\centering
	\includegraphics[width=\textwidth]{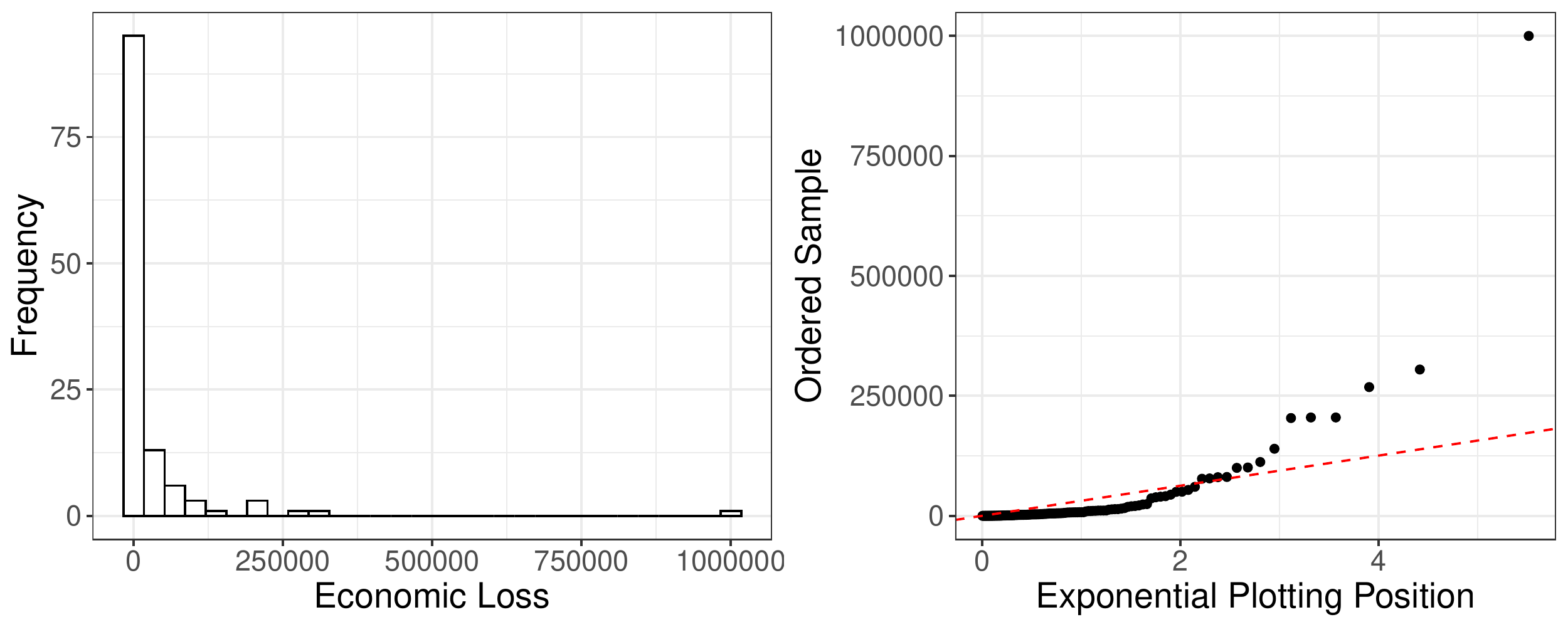}
	\caption{Histogram and exponential quantile-quantile plot of economic
losses.}\label{fig:EL_expqq}
\end{figure}

\begin{figure}
	\centering
	\includegraphics[width=\textwidth]{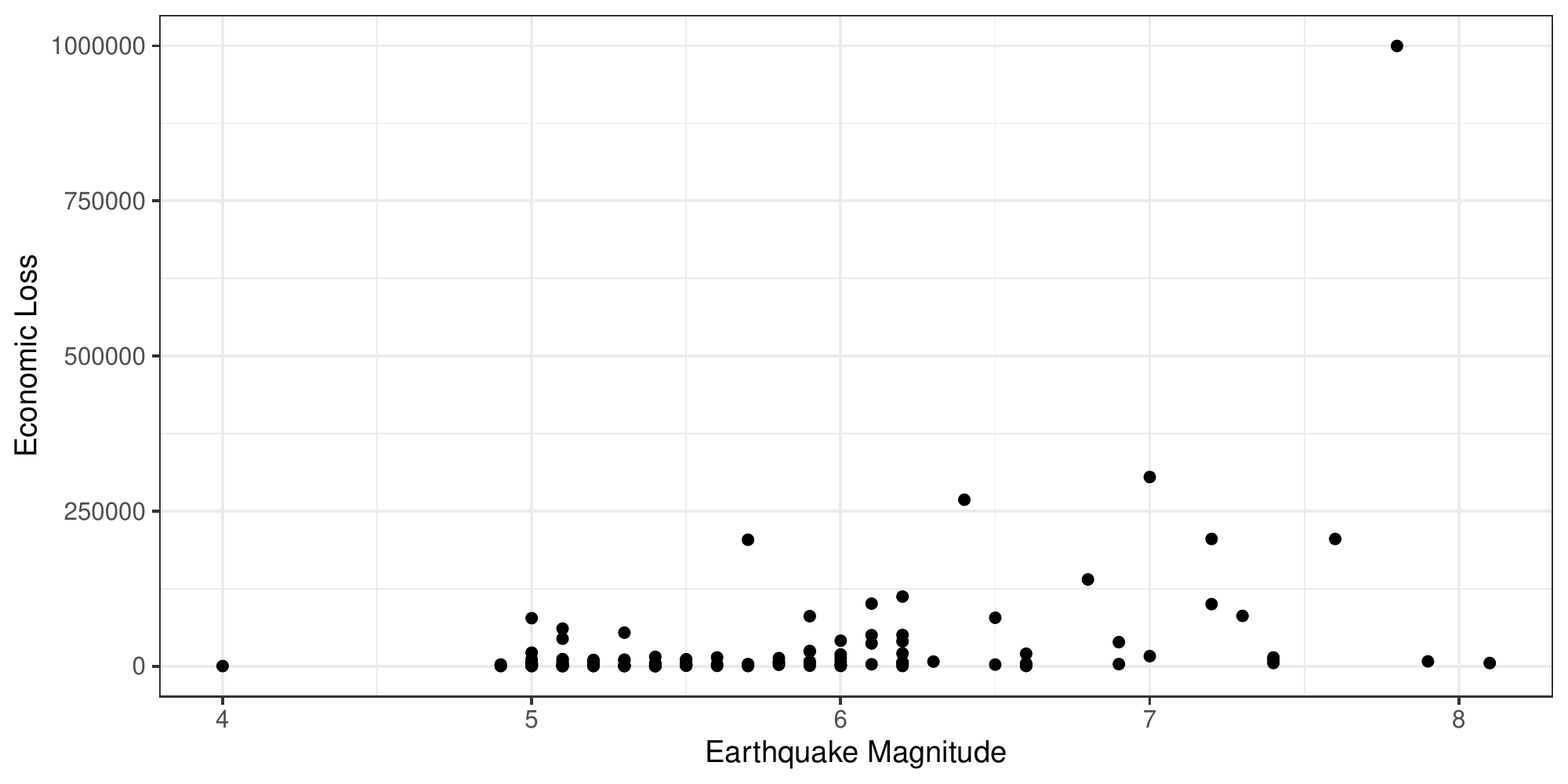}
\caption{Scatterplot of earthquake magnitudes and economic
losses.}\label{fig:scatter}
\end{figure}

In this dataset, information of 124 earthquakes are collected, among which 77
occurred in cities and 47 occurred in rural areas. A description of the dataset
is shown in Table~\ref{tab:expl}. The magnitudes of the earthquakes range from 4
to 8.1, and economic loss they incurred range from 5 CNY to $10^6$ CNY, making
an extremely wide interval. A histogram and an exponential quantile-quantile
plot of the economic losses are shown in Figure~\ref{fig:EL_expqq}, from which
we observe that the economic loss is heavily tailed. A scatterplot of economic
losses versus magnitudes is presented in Figure~\ref{fig:scatter}. It is rather
clear that simple linear regression cannot capture the relation between the
earthquake magnitudes and the economic losses.

\begin{table}[tbp]
	\center
\caption{Summary of response and covariates in the earthquake
dataset.}\label{tab:expl}
	\begin{tabular}{ccccc}
\toprule
Notation & Variable Name & Type & Range/Categories & Median/Count \\
\midrule
		$Z$ & Economic Loss & Numerical & [5, $10^6$] & 5000\\
		$X_1$ & Earthquake Magnitude& Numerical & [4, 8.1] & 5.5 \\
		$X_2$ & Urban Indicator & Binary & $\{0, 1\}$ & $\{47(X_2=0), 77(X_2=1)\}$\\
\bottomrule
	\end{tabular}
\end{table}

\section{Methodology}\label{sec:method}

\subsection{The Bayesian Hierarchical Model}

Heavy-tailed distributions are probability distributions whose tails are not
exponentially bounded, and there are even super-exponential distributions. They
have been used in many areas, such as earth science, survival analysis,
economics, and finance. Specifically, in economics or finance study
\citep{rachev2003handbook}, the underlying risk factors are
always assumed to follow
heavy-tailed distributions.
The Weibull distribution is one of the most important heavy-tailed distributions
that is used as a probabilistic model of the amount of loss associated with
actuarial and financial risk management \citep{gebizlioglu2011comparison}. With
this feature, the Weibull distribution is a reasonable choice for us to model
the economic losses caused by earthquakes.

The Weibull probability density function is
\begin{equation}
f(x)=\frac{k}{\lambda}\left(\frac{x}{\lambda}\right)^{k-1}\exp\left(
	-\left(\frac{x}{\lambda}\right)^{k}\right), ~~x>0,~ k>0,~ \lambda>0,
\end{equation}
where $k$ and $\lambda$ are the shape and scale parameters, respectively. In
order to obtain an efficienct conjugate form under Bayesian setting, we rewrite
the probability density function alternatively as
\begin{equation}\label{eq:newweibull}
f(x)=kb\cdot x^{k-1}\exp\left(-x^{k}b\right), ~~x>0,~ b=\lambda^{-k}>0,
~k>0.
\end{equation}
For the rest of this paper, we denote the distribution expressed
by~\eqref{eq:newweibull} as Weibull($k$,$b$).

To bring in spatial information, latent Gaussian models
\citep{gelfand2016spatial} have been conventionally used. Introducing spatial
random effect into the Weibull model, for locations $\bm{s} =
(\bm{s}_1,\ldots,\bm{s}_n)$, denote the losses as $Z(\bm{s}_1),\ldots,
Z(\bm{s}_n)$, we have
$$Z_{\bm{s}_i}\sim \mbox{Weibull}(k, \mu_i),$$
where the vector $\bm{\mu} = (\mu_1,\ldots, \mu_n)$ satisfy:
\begin{equation}
\log(\bm{\mu})=\bm{X}(\bm{s})\bm{\beta}+\bm{W}, 
\end{equation}
where $\bm{\beta}=(\beta_0,\ldots,\beta_p)^\top$, $\bm{W}$ is an $n$-dimensional
vector of  spatial random effects, $\bm{X}(\bm{s})$ is an~$n\times p$ covariate
matrix, and $\bm{s}_i\in \mathcal{R}^2$. Furthermore, it is often assumed that
\begin{align}
\bm{W} \mid \phi, \sigma_w \sim \mbox{MVN}(\bm{0}_n,\sigma_{w}^2\bm{H}(\phi)),
\end{align}
where $\bm{W}=(w(\bm{s}_1),\ldots,
w(\bm{s}_n))^\top$, $\bm{H}(\phi)$ is the $n\times n$ spatial correlation matrix
with $\phi$ being the range parameter, and MVN denotes the multivariate normal
distribution. In our paper, we use an exponential covariogram to
define~$\bm{H}(\phi)$,i.e.,
\begin{equation}
    \bm{H}(\phi) = \exp(-\mbox{dist}/\phi),
\end{equation}
where ``dist'' denotes an $n\times n$ matrix whose $(i,j)$th entry is the
distance between locations $\bm{s}_i$ and $\bm{s}_j$.

A Bayesian approach involves specifying prior distributions for unknown
parameters. Under the setting described above, the joint distribution of the
data, processes, and parameters is written as the product of the following
distributions:
\begin{equation}\label{eq:gaumodel}
\begin{split}
&\textbf{Data Model}: Z(\bm{s}_i)\mid  \bm{W},\bm{\beta},\sigma^2,\phi, 
\stackrel{\text{ind}}{\sim} \text{Weibull}(k,\mu(\bm{s}_i));
~~ i=1,\ldots, n; \forall k\in(0,+\infty)\\
&\textbf{Process Model}: \bm{W} \mid \phi, \sigma_w \sim \mbox{MVN}(\bm{0}_n,
\sigma_{w}^2\bm{H}(\phi))\\
&\textbf{Parameter Model 1}: \bm{\beta} \sim \mbox{MVN}(\bm{0}_p,
\sigma^2\bm{I}_p)\\
&\textbf{Parameter Model 2}: \log(\sigma) \sim \text{N}(0,1)\\
&\textbf{Parameter Model 3}: \log(\sigma_{w}) \sim \text{N}(0,1)\\
&\textbf{Parameter Model 4}: \phi \sim \text{DU}(a_1,b_1)
\end{split},
\end{equation}
where DU is shorthand for discrete uniform distribution. The formulation
in~\eqref{eq:gaumodel} requires tuning of Metropolis--Hasting steps
\citep{chib1995understanding} within the Gibbs sampler, as Gaussian process does
not maintain conjugacy for non-Gaussian data. To improve the computational
efficiency for Weibull regression, a conjugate prior is desired.
\citet{bradley2018computationally} proposed a multivariate log-gamma (MLG)
distribution as the conjugate prior for Poisson spatial regression model, and
established connection between multivariate log-gamma distribution and
multivariate normal distribution. The following construction demonstrates that
MLG is also an ideal prior choice for Weibull regression model because of their
conjugacy. Similar to \citet{bradley2018computationally}, we define the
$n$-dimensional random vector $\bm{\gamma}=(\gamma_1,...,\gamma_n)^\top$, which
consists of $n$ mutually independent log-gamma random variables with shape and
scale parameters organized into $n$-dimensional vectors $\bm{\alpha} \equiv
(\alpha_1,...,\alpha_n)^\top$ and $\bm{\kappa} \equiv
(\kappa_1,...,\kappa_n)^\top$, respectively. Then the $n$-dimensional random
vector $\bm{q}$ is defined as
\begin{equation}
\bm{q}=\bm{\mu}+\bm{V}\bm{\gamma},
\label{linear transformation}
\end{equation}
where $\bm{V}\in \mathcal{R}^n\times\mathcal{R}^n$ and $\bm{\mu}\in
\mathcal{R}^n $. \citet{bradley2018computationally} called $\bm{q}$ the
multivariate log-gamma random vector. The probability density function
of the random vector $\bm{q}$ can be defined as:
\begin{equation}\label{mlgpdf}
f(\bm{q}\mid \bm{c},\bm{V},\bm{\alpha},\bm{\kappa})=\frac{1}{\det
(\bm{V})}\left(\prod_{i=1}^m \frac{1}{\Gamma(\alpha_i)\kappa_i^{\alpha_i}
}\right)\exp\left[\bm{\alpha}^\top\bm{V}^{-1}(\bm{q}-\bm{\mu})-
\bm{\kappa}^{(-1)^\top}
\exp\{\bm{V}^{-1}(\bm{q}-\bm{\mu})\}\right],
\end{equation}
where ``det'' represents the determinant function. We use $\mbox{MLG}(\bm{\mu},
\bm{V}, \bm{\alpha},\bm{\kappa})$ as a shorthand for the probability density
function in \eqref{mlgpdf}. From \citet{bradley2018computationally}, we know
that the latent multivariate log-gamma process is a saturated process of the
latent Gaussian process. If $\bm{q}$ follows a multivariate log-gamma
distribution
$\text{MLG}(\bm{0},\alpha^{1/2}\bm{V},\alpha\bm{1},1/\alpha\bm{1})$, as $\alpha
\rightarrow \infty$, $\bm{\beta}$ will converge in distribution to a
multivariate normal distribution vector with mean $\bm{0}$ and covariance matrix
$\bm{V}\bm{V}^\top$. In practice, choosing $\alpha=10~000$ is sufficient for
this normal approximation.

This property of the MLG distribution makes it a favorable choice in our
scenario. Substituting the Gaussian distribution in \eqref{eq:gaumodel} with
MLG, we have the following hierarchical model:
\begin{equation}\label{eq:finalmodel}
\begin{split}
&\textbf{Data Model}: Z(\bm{s}_i)\mid \bm{W},\bm{\beta},\sigma^2,\phi, 
\stackrel{\text{ind}}{\sim} \text{Weibull}(\textit{k},\mu(\bm{s}_i));
\hspace{5pt} i=1,\ldots, n; \forall \textit{k}\in(0,+\infty)\\
&\textbf{Process Model}: \bm{W}\mid \phi, \sigma_w \sim \text{MLG}(\bm{0}_n,
\bm{\Sigma}^{1/2}_{{W}},\alpha_{{W}}\bm{1}_n,\kappa_{{W}}\bm{1}_n)\\
&\textbf{Parameter Model 1}: \bm{\beta} \sim \text{MLG}(\bm{0}_p,
\bm{\Sigma}^{1/2}_{{\beta}},\alpha_{{\beta}}\bm{1}_p,\kappa_{{\beta}}\bm{1}_p
)\\
&\textbf{Parameter Model 2}: \log(\sigma) \sim \text{N}(0,1)\\
&\textbf{Parameter Model 3}: \log(\sigma_{w}) \sim \text{N}(0,1)\\
&\textbf{Parameter Model 4}: \phi \sim \text{DU}(a_1,b_1)
\end{split},
\end{equation}
where \textit{k} denotes the shape parameter of the Weibull distribution,
$\mu(\bm{s}_i)=\bm{X}(s_{i})\bm{\beta}+\bm{W}$,
$\bm{\Sigma}_{{W}}=\sigma_{w}^2\bm{H}(\phi)$,
$\bm{\Sigma}_{{\beta}}=\sigma^2\bm{I}_p$,
$\alpha_{{W}}>0,\alpha_{{\beta}}>0,\kappa_{{W}}>0,$ and $\kappa_{{\beta}}>0$.
Based on the results of \citet{bradley2018computationally}, the full
conditionals of $\bm{\beta}$ and $\bm{W}$ will be the conditional MLG
distribution (cMLG). As there is no analytic forms for the posterior
distributions of $\sigma_w$ and $\phi$, in this work we use Metropolis--Hasting
algorithm \citep{chib1995understanding} to obtain posterior samples. Slice
sampling \citep{neal2003slice} might be an alternative approach, but we do not
discuss it here, and refer interested readers to the original text. Full
conditional distributions are presented in Appendix~A.

\subsection{Long-Tailed Property Justification}
We present theoretical justification for usage of MLG as a conjugate prior for
Weibull distribution. The Weibull distribution is ``long-tailed'' with shape
parameter greater than~0 but less than 1, which is an important subclass of
heavy-tailed distributions \citep{asmussen2003steady}. A random variable $X$ is
said to have a long right tail if for all $k>0$,
\begin{equation}
\lim_{x\rightarrow \infty}P[X>x+k\mid X>x]=1.
\label{longtail}
\end{equation}
For the Weibull distribution, the long tail probability can be written as
\begin{equation}
LP=P[Z>z+\delta\mid Z>z]=\frac{\exp(-b(z+\delta)^k)}{\exp(-bz^k)},
\label{Paretolongtail}
\end{equation}
where $0<k<1$ is the shape parameter of the Weibull distribution. Under the
spatial setting, $b(\bm{s})=\exp\{\bm{X}^\top(\bm{s})\bm{\beta}+w(\bm{s})\}$ for
any $\bm{s}\in D$. thus the log of the long-tail probability can be expressed as
\begin{equation}
\text{log}(LP)=\exp\{\bm{X}(\bm{s})^\top\bm{\beta}+w(\bm{s})\}(z^k-(z+\delta)^k
).
\label{loglongtail}
\end{equation}
For the latent Gaussian model of $w(\bm{s})$, it is rather straightforward to
find the expected value of $\text{log}(LP)$ using the moment generating function
for the normal distribution. The expected log long-tail probability under
Gaussian model is given as
\begin{equation} \label{gaussianlongtail}
E_{\mbox{G}}\{\text{log}(LP)\mid
\sigma^2_\beta,\sigma^2_w\}=\exp\left\{\frac{1}{2}
\left(\sum_{i=1}^pX^2_i(s,t)\sigma^2_\beta+\sigma^2_w\right)\right\}
(z^k-(z+\delta)^k).	
\end{equation}
For the multivariate log-gamma model, the expected log long-tail probability is
\begin{equation}
E_{\mbox{MLG}}\left\{\text{log}(LP)\mid
\sigma^2_\beta,\sigma^2_w,\alpha_w,\alpha
_\beta,
\kappa_\beta,\kappa_w\right\}=\frac{\left(\frac{\kappa_{\beta}^
{p\alpha_{\beta}}}{\Gamma(\alpha_\beta)^p}\right)\left(\frac
{\kappa_w^{\alpha_w}}{\Gamma(\alpha_w)}\right)}{\left(\frac{\kappa_\beta
^{p\alpha_\beta+\sum_{i=1}^pX_i(s,t)\sigma_\beta}}{\prod_{i=1}^p
\Gamma(\alpha_\beta+X_i(s,t)\sigma_\beta)}\right)\left(\frac
{\kappa_w^{\alpha_w+1}}{\Gamma(\alpha_w+1)}\right)}(z^k-(z+\delta)^k).
\label{mlglongtail}
\end{equation}
\noindent
Proofs for Equations \eqref{gaussianlongtail} and \eqref{mlglongtail} are
provided in Appendix B of \citet{hu2018bayesian}. A relationship between
Equations \eqref{gaussianlongtail} and \eqref{mlglongtail} is provided in
Proposition 1 below.
\begin{proposition}
	Assume that $\bm{\beta}$ and $\bm{W}$ follow the MLG distribution as defined
	in Equation \eqref{mlgpdf}. Then, we have the following,
	\begin{align*}
	&E_{MLG}\{\lim _{\alpha \rightarrow \infty}\log(LP)\mid \sigma^2_\beta 
	= \alpha \sigma_{1}^{2},\sigma^2_w = \alpha \sigma_{2}^{2},\alpha_w = 
	\alpha,\alpha_\beta = \alpha,\kappa_\beta = \alpha,\kappa_w = \alpha\}\\
	& =E_G\{\log(LP)\mid \sigma^2_\beta = \sigma_{1}^{2},\sigma^2_w = 
	\sigma_{2}^{2}\},
	\end{align*}
	where $E_{MLG}$ is the expected value with respect to the multivariate
	log-gamma distribution, and $E_{G}$ is the expected value with respect to
	the Gaussian distribution, $\sigma_{1}^{2}>0$, and $\sigma_{2}^{2}>0$.
\end{proposition}
\begin{proof}
	Pass the limit through the expectation, and apply Proposition 1 from 
	\citet{bradley2018computationally}.
\end{proof}

\subsection{Risk Measures Based on the Posterior Predictive Distribution}

While in conventional model fitting, researchers are concerned with the final
point prediction which often occurs at the mean, with catastrophes or extreme
environmental events such as earthquakes or hurricanes, risk measures different
from the mean are often used, which are often of high importance to policy
sellers in the insurance industry. The most popular among them include
value-at-risk \citep[VaR;][]{duffie1997overview}, expected shortfall
\citep[ES;][]{acerbi2002expected} and tail-value-at-risk
\citep[TVaR;][]{barges2009tvar}. VaR is a popular risk measure because of its
simplicity and easiness to be understood \citep{dowd2006after}. Given $\alpha$
between 0 and 1, VaR is defined as the $100(1-\alpha)$th percentile of the
density function of loss, and denoted as $q_\alpha$. Being an alternative
measure, ES is defined to be the negative of the expectation of the tail beyond
the VaR. Noticing that the VaR is only a numerical value and does not describe
the loss pattern in the tail beyond itself, \cite{barges2009tvar} proposed the
more general TVaR, which is defined as
\begin{equation}
	\mbox{TVaR}_\alpha = \frac{1}{1-\alpha} \int_\alpha^1 
	\mbox{VaR}_u \dd u,
\end{equation}
and captures the entire tail beyond the specified percentile rather than one
point, making it a measure for the average risk. For a new catastrophic event,
we would like to make predictions on the risk. In Bayesian analysis, the
posterior predictive distribution is the distribution of possible unobserved
values conditional on the observed values \citep{gelman2013bayesian}. Let
$p(\bm{\theta}|\bm{Z})$ denote the posterior distribution of all parameters
$\bm{\theta}$ given data $\bm{Z}$, then the posterior predictive distribution of
new data $\bm{Z}^*$ is given as
\begin{equation}
\label{eq:posterior_predictive}
p(\bm{Z}^*\mid \bm{Z})=\int p(\bm{Z}^*\mid \bm{\theta},\bm{Z})
p(\bm{\theta}\mid \bm{Z})\dd  \bm{\theta}.
\end{equation}
The VaR, ES, and TVaR based on the posterior predictive distribution of
$\bm{Z}^*$ are defined as:
\begin{align}
    \text{VaR}_\alpha(\bm{Z}^*)&=q_\alpha(\bm{Z}^*), \label{var} \\
    \text{ES}_\alpha(\bm{Z}^*)&=E[\bm{Z}^*
\mid \bm{Z}^*\geq\text{VaR}_\alpha(\bm{Z}^*)
], \label{es}\\
\text{TVaR}_\alpha(\bm{Z}^*)&=\frac{1}{1-\alpha}\int_{\alpha}^1
\text{VaR}_t(\bm{Z}^*) \dd t. \label{tvar}
\end{align}

\section{Bayesian Model Selection Criterion}\label{sec:model_section}

As seen in the model construction \eqref{eq:finalmodel}, tuning of parameter $k$
for Weibull distribution is needed. To select the most suitable model parameter,
we adapt the Bayesian model assessment criterion, logarithm of the
Pseudo-marginal likelihood \citep[LPML;][]{ibrahim2013bayesian} to our Weibull
regression with spatial random effects scenario. The LPML is defined as
\begin{align}
\label{eq:defLPML}
\text{LPML} = \sum_{i=1}^{n} \text{log}(\text{CPO}_i),
\end{align}
where $\text{CPO}_i$ is the conditional predictive ordinate (CPO) for the $i$-th
subject. CPO is calculated based on the leave-one-out-cross-validation, which
estimates the probability of observing data $y_i$ in the future after having
already observed data $y_1,\cdots,y_{i-1},y_{i+1},\cdots,y_n$. The CPO for the
$i$-th subject is defined as
\begin{align}
\text{CPO}_i=f(y_i\mid \bm{y}_{-i}) \equiv \int f(y_i\mid \bm{\theta})
\pi(\bm{\theta}\mid \bm{y}_{-i}) \dd \bm{\theta},
\label{eq:cpo_def}
\end{align}
where $\bm{y}_{-i}=\{y_1,\cdots,y_{i-1},y_{i+1}\cdots,y_n\}$, and
\begin{align}
\pi(\bm{\theta}\mid \bm{y}_{-i})=\frac{\prod_{j\neq i}f(y_j\mid \bm{\theta})
\pi(\bm{\theta})}{c(\bm{y}_{-i})},
\end{align}
where $c(\bm{y}_{-i})$ is the normalizing constant.
The $\text{CPO}_i$ in Equation~\eqref{eq:cpo_def} can be expressed as
\begin{align}
\text{CPO}_i=\frac{1}{\int\frac{1}{f(y_i\mid \bm{\theta})}\pi(\bm{\theta}
\mid \bm{y}_{-i})\dd \bm{\theta}}.
\label{CPO1}
\end{align}
Therefore, a Monte Carlo estimate of  $\text{CPO}_i$ in Equation
 \eqref{CPO1} is given by
\begin{align}
\widehat{\text{CPO}}^{-1}_i=\frac{1}{B}\sum_{b=1}^B\frac{1}{f(y_i\mid
\bm{\theta}_b)},
\label{CPO monte}
\end{align}
where $\bm{\theta}_b$ is the $b$-th MCMC sample of $\bm{\theta}$ 
from $\pi(\bm{\theta}\mid \bm{y})$.
For model \eqref{eq:finalmodel}, we have following estimation for CPO:
\begin{equation}
\label{CPO_final_model}
\widehat{\text{CPO}}_i^{-1}=\frac{1}{B}\sum_{i=1}^B\frac{1}
{f(Z(\bm{s_i})\mid \bm{\beta}_b,\bm{X}(\bm{s_i}),\hat{w}(\bm{s_i}))},
\end{equation}
where $\{\bm{\beta}_b, b=1,\cdots,B\}$ denotes a Gibbs sample of $\bm{\beta}$
from $\pi(\bm{\beta}\mid \bm{\text{Data}})$, and~$\hat{w}(\bm{s_i})$ is
posterior mean of spatial random effects on location $\bm{s_i}$. Finally, the
logarithm of the Pseudo marginal likelihood (LPML) is defined as
\begin{align}
\text{LPML}=\sum_{i=1}^{n} \text{log}(\widehat{\text{CPO}_i}).
\end{align}
In the context of model selection, we select the best model which has the
largest LPML value under CPO.

\section{Simulation Study}\label{sec:simu}

For simplicity, we base our simulation study on the spatial domain $D=[0,
3]\times[0, 3]$, and the locations $\bm{s}_i,i=1\cdots,n$ are generated
uniformly over $D$. The proposed model \eqref{eq:finalmodel} is fitted. In the
first three simulation designs, we set $\bm{\beta}=(-1,-1,-1)^\top$, and
$X(\bm{s}_i)$'s are independently generated from the standard uniform
distribution $\mbox{U}(0, 1)$. The shape parameter for Weibull distribution
is set to
be $k\in\{0.2, 0.5, 0.8\}$. The vector of spatial random effects $\bm{W}$ is
generated from
$\text{MLG}(\bm{0},\sigma_w\bm{H}^{1/2}(\phi),\alpha_w,\kappa_w)$ where
$\phi=5$, $\sigma_w=1$, $\alpha_w=1$, $\kappa_w=1$, and $\bm{H}$ is the matrix
of Euclidean distances between pairs in the $n$ generated locations. A DU(1,10)
prior is given to $\phi$ following Chapter~6 of \cite{carlin2014hierarchical}.
Statistical inference is performed using MCMC with chain length of 5000, and we
drop the first 2000 iterations as burn-in. We calculate the bias, standard
deviation (SD), mean squared error (MSE), and coverage rate (CR) for each
parameter based on
posterior mean and posterior quantiles. The simulation results are shown in
Tables~\ref{tab:k02},~\ref{tab:k05} and~\ref{tab:k08}. 
%


\begin{table}[tbp]
	\caption{Performance measures when $k=0.2$.}
	\label{tab:k02}
	\begin{center} 
		\begin{tabular}{llccccc}
			\toprule
			Setting &			parameter  & Bias & SD & MSE &CR \\
			\midrule  
			$n=100, k=0.2$& 	$\beta_1$ &0.0510 &0.4787&0.2317&0.94         \\
			&	$\beta_2$ &0.0405 &0.4757&0.2279&0.94               \\
			&	$\beta_3$ &0.0402 &0.4804&0.2324&  0.94                \\
			&	$\log(\sigma)$ &-0.0085&0.9495&0.9014& 0.96   \\
			&	$\log(\sigma_w)$ &0.0581  &0.8895&0.7945& 0.94      \\
			\midrule 
			$n=150, k=0.2$&	$\beta_1$ &-0.0263 &0.4264&0.1811&0.94           \\
			&	$\beta_2$ &0.0100 &0.4223&0.1784& 0.94         \\
			&	$\beta_3$ &-0.0196 &0.4194&0.1755&    0.94             \\
			&	$\log(\sigma)$ & -0.0397&0.9220&0.8485& 0.93  \\
			&	$\log(\sigma_w)$ &0.0422 &0.9296&0.8659& 0.94       \\
			\midrule
			$n=200, k=0.2$&	$\beta_1$ &0.0054&0.3398&0.1154& 0.94          \\
			&$\beta_2$ &0.0217  &0.3361&0.1134& 0.94        \\
			&$\beta_3$ & 0.0060 &0.3383&0.1144&  0.94              \\
			&	$\log(\sigma)$ &0.0738&0.9005&0.8163&0.91    \\
			&$\log(\sigma_w)$ & 0.0235&0.9019&0.8139&0.94        \\
			\bottomrule
		\end{tabular}
	\end{center}
\end{table}

\begin{table}[tbp]
	\caption{Performance measures when $k=0.5$.}
\label{tab:k05}
	\begin{center} 
		\begin{tabular}{llccccc}
			\toprule
Setting &			parameter  & Bias & SD & MSE &CR \\
			\midrule  
		$n=100, k=0.5$& 	$\beta_1$ &0.0327 &0.4808&0.2322& 0.94        \\
		&	$\beta_2$ &0.0253 &0.4752&0.2264&   0.94          \\
		&	$\beta_3$ &-0.0054 &0.4774&0.2278&  0.94              \\
		&	$\log(\sigma)$ &0.0156&0.9232&0.8525& 0.92   \\
		&	$\log(\sigma_w)$ &0.0058  &0.9246&0.8549&  0.94   \\
\midrule 
		$n=150, k=0.5$&	$\beta_1$ &0.0262 &0.4022&0.1624&0.94           \\
		&	$\beta_2$ &0.0023 &0.4009&0.1607&0.94          \\
		&	$\beta_3$ &0.0212 &0.4112&0.1695& 0.94                \\
		&	$\log(\sigma)$ &0.0439&0.9261&0.8595& 0.91   \\
		&	$\log(\sigma_w)$ &-0.0029 &0.9150&0.8372&  0.94      \\
		\midrule
		$n=200, k=0.5$&	$\beta_1$ &-0.0225&0.3687&0.1354& 0.94         \\
			&$\beta_2$ & -0.0003 &0.3585&0.1285&0.94        \\
			&$\beta_3$ &0.0229  &0.3645&0.1333&   0.94            \\
			&	$\log(\sigma)$ &-0.0158&0.9106&0.8289& 0.96   \\
			&$\log(\sigma_w)$ &0.0412 &0.9264&0.8599& 0.94      \\
			\bottomrule
		\end{tabular}
	\end{center}
\end{table}

\begin{table}[tbp]
	\caption{Performance measures when $k=0.8$.}
\label{tab:k08}
	\begin{center} 
		\begin{tabular}{llccccc}
			\toprule
Setting &			parameter  &  Bias & SD & MSE &CR \\
			\midrule  
		$n=100, k=0.8$& $\beta_1$ &0.0021&0.5222&0.2726&  0.94       \\
		&	$\beta_2$ &0.0326&0.5365&0.2888&0.94 \\
		&	$\beta_3$ &0.0654&0.5314&0.2866&0.94 \\
		&	$\log(\sigma)$ &-0.0159 &0.9106&0.8289& 0.95\\
		&	$\log(\sigma_w)$ &0.0255&0.9143&0.8365& 0.94 \\
\midrule 
		$n=150, k=0.8$&	$\beta_1$ &0.0474  & 0.4054&0.1665&  0.94     \\
		&	$\beta_2$ &-0.0475 &0.4055&0.1621& 0.94\\
		&	$\beta_3$ &-0.0096 &0.4058& 0.1645&  0.94      \\
		&	$\log(\sigma)$ &0.0368 &0.9156&0.8396& 0.94  \\
		&	$\log(\sigma_w)$ &-0.0664 &0.9039 &0.8126& 0.94   \\
		\midrule
		$n=200, k=0.8$&	$\beta_1$ &0.0350  &0.3657  &0.1349&0.94      \\
			&$\beta_2$ & -0.0296 &0.3672 &0.1339&0.94   \\
			&$\beta_3$ &-0.0281  &0.3539&0.1244& 0.94       \\
			&$\log(\sigma)$ &-0.0381 &0.9248&0.8538& 0.93  \\
			&$\log(\sigma_w)$ &-0.0179 &0.9312 &0.8668&0.94   \\
			\bottomrule
		\end{tabular}
	\end{center}
\end{table}


A few observations can be made from Tables~\ref{tab:k02},~\ref{tab:k05}
and~\ref{tab:k08}. For each value of $k$, even with sample size of 100, the
model parameters are estimated with very small average bias. The bias,
SD and MSE all decrease when the number of observations increase. The CP
remains close to its 0.95 nominal level. Comparing across tables, it can be
seen that as the shape parameter $k$ increases, the SD and MSE of parameter
estimates increase.

To demonstrate the performance of our proposed model under a different scenario
where the underlying spatial random effects are generated from multivariate
normal distribution, in the last simulation scenario for each $k$, we
generate~$\bm{W}$ from
multivariate normal distribution with mean zero and covariance
matrix~$\bm{\Sigma}_{{W}}=\sigma^2_w\bm{H}(\phi)$. We assume true~$\phi=5$
and~$\sigma_w=1$. The covariate~$X(\bm{s}_i)$'s are independently generated
from the standard uniform distribution $\mbox{U}(0, 1)$. We again fit the model
in \eqref{eq:finalmodel} over 100 replicates. In each replicate, we
run 5000 iterations and drop the first 2000
iterations as burn-in. The estimation performance results are shown in
Table~\ref{tab:MVN}. 
The biases of parameter estimates increased as a consequence of model
misspecification. In all three scenarios, the parameter estimates are biased
to the positive direction, which indicates the effect of covariates are
underestimated as the MLG is long-tailed, and its long-tail captured more
information than needed. The SD, however, is smaller when compared to the
corresponding scenarios when~$\bm{W}$ is generated from MLG. The CR still remain
close to its nominal level. In addition, the estimates for $\log(\sigma)$ and
$\log(\sigma_w)$ remain very precise even if the model is misspecified.

\begin{table}[tbp]
\caption{Performance measures when spatial random effect generate from
multivariate normal distribution.}
	\label{tab:MVN}
	\begin{center} 
		\begin{tabular}{llccccc}
			\toprule
			Setting &			parameter  &  Bias & SD & MSE &CR \\
			\midrule  
			$n=200, k=0.2$& $\beta_1$ &0.2055&0.3058&0.1357&0.94     \\
			&	$\beta_2$ &0.0906&0.3035&0.1003&0.94 \\
			&	$\beta_3$ &0.1066&0.3061&0.1051&0.94 \\
			&	$\log(\sigma)$ &-0.0105&0.9389&0.8814&0.97\\
			&	$\log(\sigma_w)$ &-0.0103&0.9267&0.8586&0.94 \\
			\midrule 
			$n=200, k=0.5$&	$\beta_1$ &0.1722&0.3053&0.1229&0.94   \\
			&	$\beta_2$ &0.1987&0.3052&0.1326&0.94\\
			&	$\beta_3$ &0.1703&0.3039&0.1214& 0.94  \\
			&	$\log(\sigma)$ &0.0260&0.9109&0.8304&0.94 \\
			&	$\log(\sigma_w)$ &-0.0188&0.9140&0.8350&0.94  \\
			\midrule
			$n=200, k=0.8$&	$\beta_1$ &0.3828&0.3025&0.2380&0.94      \\
			&$\beta_2$ & 0.4114&0.2976&0.2578&0.94  \\
			&$\beta_3$ &0.3887&0.3003&0.2413& 0.94     \\
			&$\log(\sigma)$ &-0.0005&0.9346&0.8735&0.95  \\
			&$\log(\sigma_w)$ &0.0237&0.9055&0.8205&0.94    \\
			\bottomrule
		\end{tabular}
	\end{center}
\end{table}

\section{A Real Data Example}\label{sec:real_date}

We analyze the earthquake data, which includes 124 earthquakes occurred at
different locations and corresponding direct economic losses in Yunnan Province
caused by these earthquakes. In this dataset we have three covariates: a
continuous variable magnitude characterizing the relative size of an earthquake,
a binary variable county indicating whether the earthquake occurs in city or
rural area, and another binary variable indicating whether the earthquake
location is in Yunnan or not. The model in \eqref{eq:finalmodel} is considered,
with $\alpha_{\beta}=10~000$ and $\kappa_{\beta}=0.0001$, which are the values
that lead to an MLG that approximates a multivariate normal distribution. The
full conditionals in Appendix~A are used to run a Gibbs sampler. The number of
iterations of the Gibbs sampler is 25~000, and we drop the first 20~000
iterations as burn-in.

We fit Weibull regression models with nine different shape parameter values
varying from 0.1 to 0.9, and present the corresponding LPML values of those
models in Table~\ref{realdata}. The model with $k=0.7$ turned out to have the
largest LPML, and is selected as the best model. The posterior mode for $\phi$
is 1, indicating moderate spatial dependency. Table~\ref{realdataestimate} shows
the posterior estimation result of the selected model. We see that the 95\%
highest posterior density intervals \citep[HPD;][]{chen1999monte} of $\beta_1$
and $\beta_2$ does not contain zero, which means that both covariates magnitude
and county are significant. The HPD interval for $\beta_3$ contains 0, which
indicates that Yunnan province may still suffer from huge economic loss even due
to an earthquake that happened somewhere else. Since the posterior estimations
of both $\beta_1$ and $\beta_2$ are negative, we conclude that (i) earthquakes
with higher magnitudes cause larger economic losses; (ii) earthquakes occurring
in city area cause larger economic losses.

\begin{table}[t]
\caption{LPML Values of Candidate Models.}
\label{realdata}
	\begin{center} 
		\begin{tabular}{cccc}
			\toprule
			Shape      &   LPML&Shape&LPML \\
			\midrule  
			$k=0.1$&-1464 &$k=0.6$&-1288\\
			$k=0.2$&-1378 &$k=0.7$&-1263  \\
			$k=0.3$&-1333 &$k=0.8$&-1287  \\
			$k=0.4$&-1298 &$k=0.9$&-1409  \\
			$k=0.5$&-1277 &&  \\
			\bottomrule
		\end{tabular}
	\end{center}
\end{table}

\begin{table}[t]
\caption{Posterior Estimation Based on Best Model ($k=0.7$)} 
\label{realdataestimate}
	\begin{center} 
		\begin{tabular}{lccc}
			\toprule
			&  Posterior Mean& Standard Error & HPD
			Interval\\
			\midrule  
			$\beta_1$  &-0.889&0.047&(-0.977, -0.792 )\\
			$\beta_2$  &-1.345 &0.313&(-1.928, -0.710 )\\
			$\beta_3$  &-0.508 &0.353&(-1.224, 0.165)\\
			$\log(\sigma_w)$ &-0.582 &1.026 &(-2.539, 1.343) \\
			\bottomrule
		\end{tabular}
	\end{center}
\end{table}

\begin{table}[t]
	\caption{Risk Measure calculated based on the final model selected
	by LPML (Unit: CNY)}

	\label{riskmeasure}
	\begin{center} 
		\begin{tabular}{crrr}
			\toprule
			&   VaR&ES&TVaR \\
			\midrule  
			90\% & 418,444. & 4,655,642 & 4,350,369\\
			95\% & 1,833,555 & 8,016,257 & 7,192,600\\
			99\% & 12,402,260 & 16,740,000 & 14,591,333\\
			\bottomrule
		\end{tabular}
	\end{center}
\end{table}

The three risk measurements VaR, ES and TVaR are computed based on posterior
predictive distribution of the selected model under three different confidence
levels. The measures are shown in Table~\ref{riskmeasure}. They together can
help insurance companies or the government make further decisions about
catastrophic insurance coverages, reinsurance levels, or catastrophic reserves.
$\text{VaR}_{90\%}(\bm{Z}) = 418,444$ implies that the insurance coverage should
be 418,444 CNY to cover 90\% loss resulted from earthquakes. $\text{ES}_{90\%}
(\bm{Z}) =4,655,642$ means that the expected excess of loss beyond 418,444 CNY
is 4,655,642 CNY, which can be used to calculate the reinsurance premium when
considering an excess of loss reinsurance. $\text{TVaR}_{90\%} (\bm{Z})
=4,350,369$ indicates that the mean loss above 418,444 is 4,350,369 CNY.

\section{Discussion}\label{sec:discussion}

In this article, we propose an efficient Bayesian spatial model to analyze
extreme losses caused by catastrophes. Our main methodological contribution is
to use multivariate log-gamma process model for both regression coefficients and
spatial random effects within a hierarchical spatial regression model.
Multivariate log-gamma process models have the computational advantage of being
conjugate with the Weibull likelihood, and therefore allow by-pass of tuning for
Metropolis--Hastings algorithms. Additionally, our simulation results indicate
that multivariate log-gamma process models have good estimation performance even
when the data are generated from Gaussian process model. The results in this
article can be applied to analyze the losses caused by many different perils
(hurricane, tornado, earthquake, wildfire, etc.), and thus the methodology is of
independent interest.

Three topics beyond the scope of this paper are worth further investigation.
In this work we considered four covariates. When the number of covariates
becomes large, introducing Bayesian variable selection to the MLG model becomes
a necessary procedure.
In this work, we considered using MLG as the prior for Weibull distribution,
which belongs to the family of generalized extreme value (GEV) distributions.
Using the MLG as prior for other members of GEV family, such as Gumbel and
Fr\'echet distributions, are also of research interest. Finally, in our model
formulation, the spatial random effect is dependent only on distances between
pairs of locations. There could be other factors that control the spatial
random effect, such as similarities in infrastructure, and incorporating random
neighborhood structures similar to \cite{gao2019bayesian} in spatial extreme
value modeling is devoted for future research.

\section*{Acknowledgment}
Dr. Hu's research was supported by Dean's office of the College of Liberal
Arts
and Sciences at University of Connecticut.

\section*{Appendix A: Full Conditionals Distributions for
Weibull Data with Latent Multivariate Log-gamma Process Models}
From the hierarchical model in \eqref{eq:finalmodel}
, the full conditional distribution for $\bm{\beta}$ satisfies
\begin{equation}
\begin{split}
f(\bm{\beta}\mid \cdot )&\propto f(\bm{\beta})\prod f(Z\mid \cdot)\\
&\propto \exp\left[\sum_i (\bm{X}(s_i)^\top\bm{\beta}+w(s_i))-\sum_i
(Z_i^k\exp((\bm{X}(s_i)^\top\bm{\beta}+w(s_i))))\right]\\
&\times \exp\left\{\alpha_{\beta}\bm{1}^\top_p\Sigma_{\bm{\beta}}^{-1/2}
\bm{\beta}-{\kappa_{\beta}}\bm{1}^\top_p\exp(\Sigma_{\bm{\beta}}^{-1/2}\bm
{\beta})\right\}.
\end{split}	
\end{equation}
Rearranging the terms we have 
\begin{equation}
f(\bm{\beta}\mid \cdot )\propto \exp\left\{
\bm{\alpha}_{{\beta}}^\top\bm{H}_{{\beta}}\bm{\beta}-\bm{\kappa}_{{\beta}}^\top
\exp(\bm{H}_{{\beta}}\bm{\beta})\right\},
\end{equation}
which implies that $f(\bm{\beta}|\cdot )$ is equal to
$\text{cMLG}(\bm{H}_{{\beta}},\bm{\alpha}_{{\beta}},\bm{\kappa}_{{\beta}})$,
where ``$\text{cMLG}$'' is the conditional multivariate log gamma distribution
from \citet{bradley2018computationally}.

Similarly, the full conditional distribution for $\bm{W}$ satisfies
\begin{equation}
\begin{split}
f(\bm{W}\mid \cdot )&\propto f(\bm{W})\prod f(Z\mid \cdot)\\
&\propto \exp\left[\sum_i (\bm{X}(s_i)^\top\bm{\beta}+w(s_i))-\sum_i (Z_i^k
\exp((\bm{X}(s_i)^\top \bm{\beta}+w(s_i))))\right]\\
&\times \exp\left\{\alpha_W\bm{1}^\top_n\Sigma_{W}^{-1/2}\bm{W}
-{\kappa_{W}}\bm{1}^\top_n\exp(\Sigma_{W}^{-1/2}\bm{W})\right\}.
\end{split}	
\end{equation}
Rearranging the terms we have 
\begin{equation}
f(\bm{W} \mid \cdot )\propto \exp\left\{\bm{\alpha}_{{W}}^\top\bm{H}_{{W}}
\bm{W}-\bm{\kappa}_{{W}}^\top\exp(\bm{H}_{{W}}\bm{W})\right\},
\end{equation}
which implies that $f(\bm{W}\mid \cdot )$ is equal to $\text{cMLG}
(\bm{H}_{{W}},\bm{\alpha}_{{W}},\bm{\kappa}_{{W}})$.
Thus we obtain the following full-conditional distributions to
 be used within a Gibbs sampler:
\begin{align}
\begin{split}
\bm{\beta} &\sim \text{cMLG}(\bm{H}_\beta,
\bm{\alpha}_\beta,\bm{\kappa}_\beta)\\
\bm{W} &\sim \text{cMLG}(\bm{H}_W,\bm{\alpha}_W,\bm{\kappa}_W)\\
\log(\sigma) &\propto \text{MLG}(\bm{0},\bm{\Sigma}^{1/2}_{{\beta}},
\alpha_{{\beta}}\bm{1}_p,\kappa_{{\beta}}\bm{1}_p)\times\text{N}(0,1)\\
\log(\sigma_w) &\propto \text{MLG}(\bm{0},\bm{\Sigma}^{1/2}_{{W}},
\alpha_{{w}}\bm{1}_n,\kappa_{{w}}\bm{1}_n)\times\text{N}(0,1)\\
\phi  &\propto \text{MLG}(\bm{0}_n,\bm{\Sigma}^{1/2}_{{W}},
\alpha_{{w}}\bm{1}_n,\kappa_{{w}}\bm{1}_n)\times \text{DU}(a_1,b_1)
\end{split}	
\label{fullcondition2}
\end{align}
A motivating feature of this conjugate structure is that it is relatively
straightforward to simulate from a $\text{cMLG}$. For $\log(\sigma)$,
$\log(\sigma_w)$ and $\phi$, we consider using a Metropolis-Hasting algorithm
or
slice sampling procedure.
The parameters of the conditional multivariate log-gamma distribution are
summarized in Table~\ref{mlgposterior}.
\begin{table}[tbp]
	\caption{Parameters of the full conditional distribution}
	\label{mlgposterior}
	\center
	\begin{tabular}{ll}
\toprule
		Parameter&Form\\
\midrule
		$\bm{H}_\beta$& $\begin{bmatrix}
		\bm{X}\\
		\Sigma_{\bm{\beta}}^{-1/2}	
		\end{bmatrix}$
		\\
		$\bm{\alpha}_\beta$& $\begin{bmatrix}
		\bm{1}_{n\times 1}
		\\
		\alpha_{\beta}\bm{1}_{p\times 1}\end{bmatrix}$\\
		$\bm{\kappa}_\beta$&$\begin{bmatrix}
		\\{\textbf{Z}^\top}^k\bm{1}_{n\times 1}\exp(\textbf{W}^\top))
		\\
		\kappa_{\beta}\bm{1}^\top_p \end{bmatrix}$
		\\
		$\bm{H}_W$& $\begin{bmatrix}
		I_n\\
		\bm{\Sigma}^{-1/2}_{{W}}	
		\end{bmatrix}$
		\\
		$\bm{\alpha}_W$& $\begin{bmatrix}
		\bm{1}_{n\times 1}\\
		\alpha_{W}\bm{1}_{n\times 1}
		\end{bmatrix}$\\
		$\bm{\kappa}_W$&$\begin{bmatrix}
		{\textbf{Z}^\top}^k\bm{1}_{n\times 1}\exp((\bm{X}\bm{\beta})^\top)\\
		\kappa_W\bm{1}^\top_n \end{bmatrix}$ \\
\bottomrule
	\end{tabular}
\end{table} 

%
%


\end{document}